\documentclass[reprint, prl, aps, twocolumn, showkeys]{revtex4-1}

\usepackage{amssymb}
\usepackage{amsmath}
\usepackage{graphicx}
\usepackage{bm}
\usepackage{dcolumn}
\usepackage{color}
\usepackage{ulem}
\usepackage{float}
\usepackage{textcomp}
\restylefloat{table}
\usepackage{verbatim}
\usepackage{epstopdf}
\usepackage[colorlinks]{hyperref}
\usepackage[mediumspace,mediumqspace,squaren]{SIunits}

\begin{document}

\title{Unconventional anisotropic even-denominator fractional quantum Hall state in a system with mass anisotropy}
\date{\today}

\author{Md.\ Shafayat Hossain}
\author{Meng K.\ Ma}
\author{Y. J.\ Chung}
\author{L. N.\ Pfeiffer} 
\author{K. W.\ West}
\author{K. W.\ Baldwin}
\author{M.\ Shayegan}
\affiliation{Department of Electrical Engineering, Princeton University, Princeton, New Jersey 08544, USA}


\begin{abstract}

The fractional quantum Hall state (FQHS) observed at a half-filled Landau level in an interacting two-dimensional electron system (2DES) is among the most exotic states of matter as its quasiparticles are expected to be Majoranas with non-Abelian statistics. We demonstrate here the unexpected presence of such a state in a novel 2DES with a strong band-mass anisotropy. The FQHS we observe has unusual characteristics. While its Hall resistance is well-quantized at low temperatures, it exhibits highly-anisotropic in-plane transport resembling compressible stripe/nematic charge-density-wave phases. More striking, the anisotropy sets in suddenly below a critical temperature, suggesting a finite-temperature phase transition. Our observations highlight how anisotropy modifies the many-body phases of a 2DES, and should further fuel the discussion surrounding the enigmatic even-denominator FQHS.

\end{abstract}

\maketitle 

A hallmark of a high-quality, interacting two-dimensional electron system (2DES) is the mysterious even-denominator fractional quantum Hall state (FQHS) \cite{Willett.PRL.1987, Suen.PRL.1992, Manoharan.PRB.1994, Willett.RPP.2013, Falson.Nat.Phys.2015, Li.Science.2017, Zibrov.Nature.2017}, which is expected to harbor Majorana excitations obeying non-Abelian statistics and be of potential use in topological quantum computing 
\cite{Moore.Nucl.Phys.B.1991, Greiter.PRB.1992, Nayak.Rev.Mod.Phys.2008, Zhu.PRB.2016}. Here we expose some remarkable properties of this state in a 2DES, confined to an AlAs quantum well, where the electrons possess an anisotropic in-plane effective mass. When the first-excited ($N=1$) Landau level (LL) is half filled, the ground state 
has highly-anisotropic longitudinal resistances, very much reminiscent of the compressible stripe phases \cite{Koulakov.PRL.1996, Fogler.PRB.1996, Moessner.PRB.1996, Fradkin.PRB.1999, Fradkin.PRL.2000, Lilly1.PRL.1999, Du.SSC.1999, Fradkin.ARCMP.2010} reported in GaAs 2DESs for the $N>1$ LLs but, surprisingly, the Hall resistance is well quantized at the half-integer value. Remarkably, the 2DES appears to make a sudden transition to this incompressible, anisotropic phase below a critical temperature. In addition, the quantized Hall plateau is robust even at very large in-plane magnetic fields. The surprises continue away from the half filling, where another state resembling bubble phases \cite{Lilly1.PRL.1999, Du.SSC.1999, Eisenstein.PRL.2002} emerges but its Hall resistance approaches an unexpected quantized value.

We performed experiments on a 2DES confined to a 20-nm-wide, AlAs quantum well where two conduction-band minima (or valleys) centered at the edges of the Brillouin zone are occupied \cite{Shayegan.AlAs.Review.2006}. Each of these valleys, which we denote by X and Y, has an elliptical Fermi contour with anisotropic effective masses $m_l=1.0$ and $m_t=0.20$ (in units of free electron mass) along their principal longitudinal and transverse directions; X has its longitudinal axis along the [100] crystal direction, and Y along [010] (see Fig. 1(a)). Our sample has a density of $n\simeq3.2\times10^{11}$ cm$^{-2}$ and a mobility of $\simeq1.4 \times10^{6}$ cm$^{2}$/Vs which is extremely high, considering the very large electron effective mass in the system \cite{Chung.PRM.2018}. We tune the valley occupancy via the application of uniaxial strain ($\varepsilon$); see Fig. 1(a) \cite{Shayegan.AlAs.Review.2006, Suppl.Mat.}. For $\varepsilon>0$, electrons are transferred from X to Y while the reverse happens for $\varepsilon>0$ \cite{Shayegan.AlAs.Review.2006}.

\begin{figure}[b!]
\includegraphics[width=.5\textwidth]{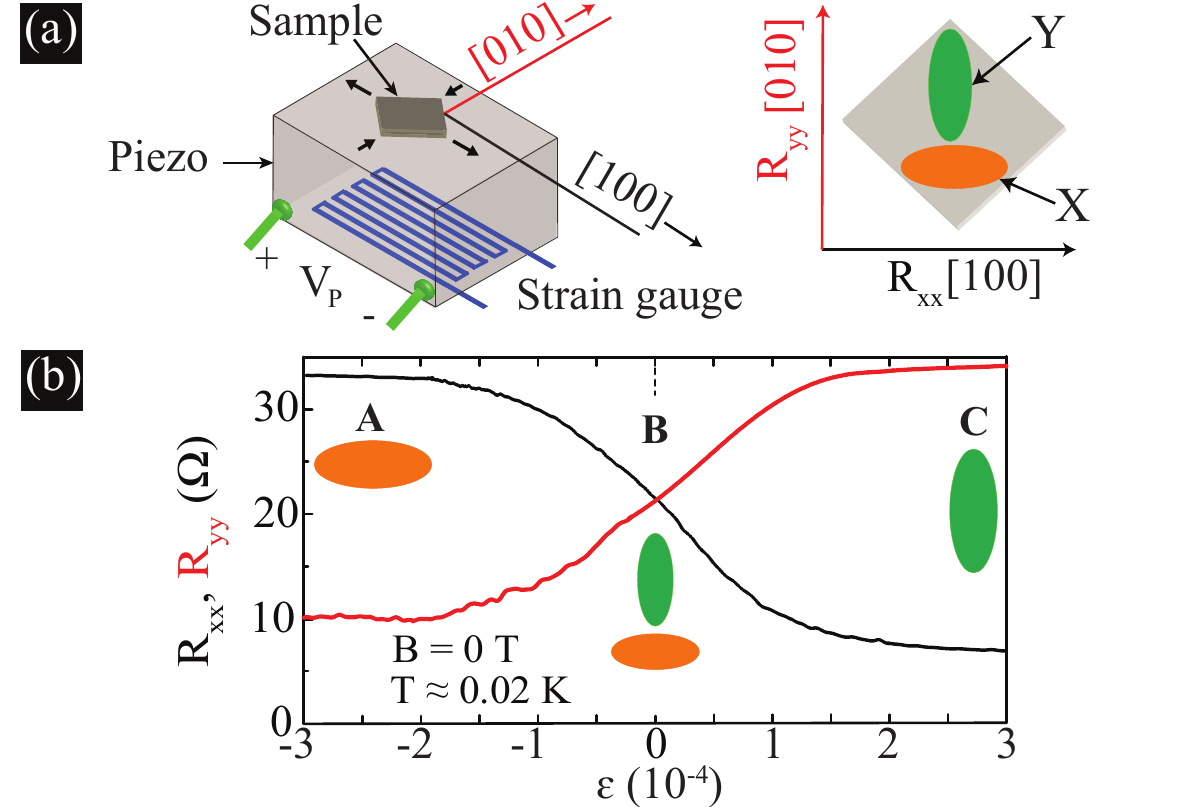}
\caption{\label{fig:Fig1} 
(a) Schematic of the experimental setup for applying in-plane strain ($\varepsilon$). The sample and a strain gauge are glued to the opposite sides of a piezo-actuator, and strain is introduced when a bias voltage ($V_{P}$) is applied to the actuator's leads. The sample geometry, including the orientation of the two occupied valleys (X and Y) and the measured resistances ($R_{xx}$ and $R_{yy}$) are shown on the right; [100] and [010] refer to the crystallographic directions. (b) Piezo-resistance of the sample as a function of $\varepsilon$. Points \textbf{A} - \textbf{C} mark the valley occupation.} 
\end{figure}

\begin{figure*}[t!]
\includegraphics[width=.99\textwidth]{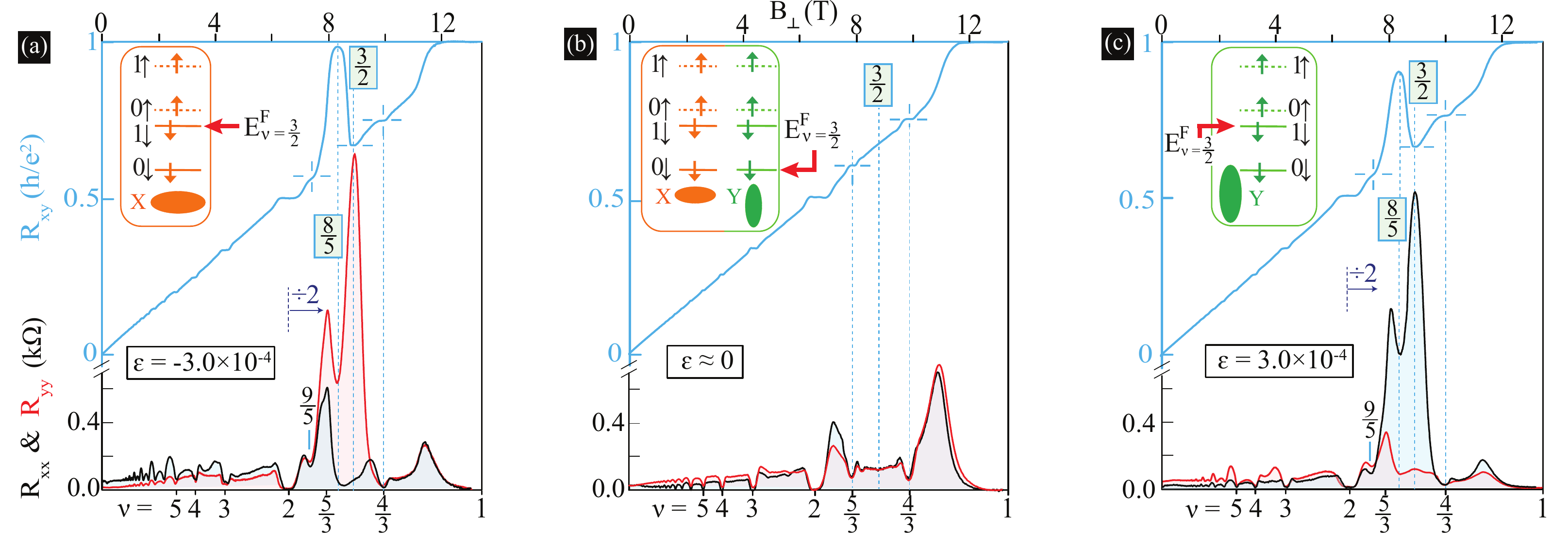}
\caption{\label{fig:Fig2} 
Exotic phases in the $N=1$ LL. Plotted are the magneto-resistance traces, $R_{xx}$ (black), $R_{yy}$ (Red) and $R_{xy}$ (blue), taken at $T\simeq 0.02$ K and at three strain values as indicated. Insets show the LLs for X (orange) and Y (green) valleys; the validity of these LL diagrams is discussed in the Supplemental Material \cite{Suppl.Mat.}. In (a) only X is occupied and $E^F_{3/2}$ resides in the X1$\downarrow$ LL, while in (c) only Y is occupied and $E^F_{3/2}$ lies in the Y1$\downarrow$ LL. In both cases $R_{xx}$ and $R_{yy}$ are highly anisotropic near $\nu=3/2$, with the hard axis along the low-mass (transverse) direction of the occupied valley, but there is an $R_{xy}$ Hall plateau at $\nu = 3/2$ quantized at $(2/3)(h/e^2)$. Transport is also highly anisotropic near $\nu=8/5$, with an $R_{xy}$ that approaches $h/e^2$. None of these anomalies are seen in (b) where $E^F$ lies in an $N=0$ LL.}
\end{figure*}

Figure 1(b), which shows the sample's piezo-resistance as a function of $\varepsilon$ at zero magnetic field, demonstrates how we tune and monitor the valley occupancy. At $\varepsilon=0$ (point \textbf{B}), because X and Y are equally populated, the 2DES exhibits isotropic transport, namely, the resistances measured along the [100] and [010] directions ($R_{xx}$ and $R_{yy}$, respectively) are equal, even though the individual valleys are anisotropic. For $\varepsilon>0$, as electrons transfer from X to Y, $R_{xx}$ decreases (black trace in Fig. 1(b)) because the electrons in Y have a small effective mass and therefore higher mobility along [100] \cite{Shayegan.AlAs.Review.2006}. (Note that the total 2DES density remains fixed as strain is applied.) The resistance eventually saturates (point \textbf{C}), when all electrons are in the Y valley \cite{Shayegan.AlAs.Review.2006, Gokmen.Natphy.2010}. For $\varepsilon<0$, $R_{xx}$ increases and eventually saturates (point \textbf{A}) as the electrons are transferred to X which has a large mass and a low mobility along [100]. As expected, the behavior of $R_{yy}$ is opposite to that of $R_{xx}$.

In Fig. 2 we show $R_{xx}$, $R_{yy}$, and $R_{xy}$ traces taken as a function of perpendicular magnetic field ($B_{\perp}$) at three values of strain, corresponding to points \textbf{A}, \textbf{B} and \textbf{C} in Fig. 1(b). The lower scales show the LL filling factor defined as $\nu=hn/eB$. The most striking features are seen in the range $2>\nu>1$. In particular, near $\nu=3/2$ the 2DES becomes strongly anisotropic at large magnitudes of strain when the electrons occupy the X (left panel) or Y (right panel) valleys. The LL diagrams shown in Fig. 2 help unravel the origin of these unusual features. Here the LLs are indicted by their orbital index ($N=0,1$) followed by the spin orientation ($\downarrow$ and $\uparrow$). Note that, thanks to the large spin-susceptibility of our 2DES, the Zeeman energy is $\simeq1.35$ times the cyclotron energy, as these LL diagrams indicate \cite{Suppl.Mat.}. When both valleys are occupied (panel (B)) the Fermi energy at $\nu=3/2$ ($E^F_{3/2}$) lies in the 0$\downarrow$ LL of X and Y. As expected for an $N=0$ LL, there are well-developed FQHSs at odd-denominator fillings $\nu=5/3$, and $4/3$, with hints of a developing FQHS at $8/5$, and transport is nearly isotropic, i.e., $R_{xx}\simeq R_{yy}$ \cite{Suppl.Mat.}. However, for single-valley occupancy, $E^F_{3/2}$ moves to the 1$\downarrow$ LL of X (left panel) or Y (right panel), and leads to the dramatic resistance anisotropy near $\nu=3/2$, which is the highlight of our work.

\begin{figure*}[t!]
\includegraphics[width=.99\textwidth]{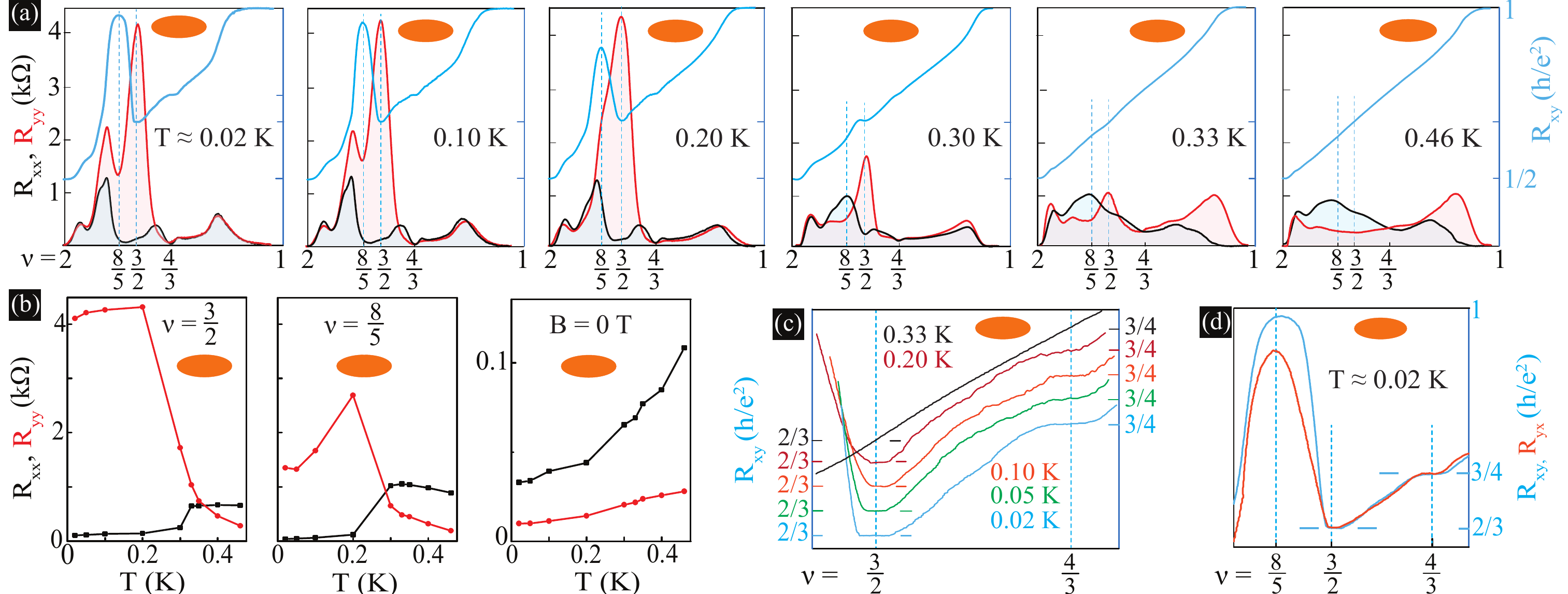}
\caption{\label{fig:Fig3} 
Temperature dependence of data for $N=1$ LL taken at $\varepsilon =-3 \times 10^{-4}$ where the 2D electrons occupy only the X valley. (a) Plots of $R_{xx}$, $R_{yy}$, and $R_{xy}$ vs filling for six different temperatures ranging from $\simeq$ 0.02 to 0.46 K. (b) Evolution of $R_{xx}$ and $R_{yy}$ at $\nu =3/2$, $8/5$, and $B=0$ with temperature. (c) $T$-dependence of $R_{xy}$, near Hall plateaus at $\nu=3/2$ and $4/3$. The traces are shifted vertically for clarity, and the horizontal lines mark the expected positions of the Hall plateaus at $\nu =3/2$ and $4/3$. (d) $R_{xy}$ and $R_{yx}$, the latter taken after switching the current and voltage leads.
}
\end{figure*}

Focusing on data at $\nu=3/2$, magneto-transport becomes highly anisotropic, exhibiting $R_{xx}<<R_{yy}$ when electrons occupy X and $R_{xx}>>R_{yy}$ for Y. Note that this anisotropy is opposite to the anisotropy seen at $B=0$ in Fig. 1(b). The anisotropic behavior in Figs. 2(a) and 2(c) is highly reminiscent of data in high-quality GaAs 2DESs in the higher LLs ($N\geq2$) in a purely perpendicular field, and generally interpreted to signal stripe phases \cite{Lilly1.PRL.1999, Du.SSC.1999}. Here we observe such anisotropy in the $N=1$ LL where normally an isotropic FQHS is seen \cite{Willett.PRL.1987, Willett.RPP.2013, Lilly1.PRL.1999, Du.SSC.1999, Eisenstein.PRL.2002}. At first sight, our data are consistent with recent theories \cite{Yang.PRB.2012, Zhu.PRB.2017} which predict that, unlike the $N=0$ LL \cite{Suppl.Mat.}, the incompressible FQHSs in the $N=1$ LL should give way to compressible stripe phases \cite{Zhu.PRB.2017}. Similar to the stripe phases in the higher ($N\geq2$) LLs, such phases are expected to consist of stripes with fillings of the two nearby integer QHSs \cite{Lilly1.PRL.1999, Du.SSC.1999}. In fact, Ref. \cite{Zhu.PRB.2017} concludes that the stripes should orient themselves along the direction of the larger mass, rendering this direction as the ``easy axis,'' consistent with our observation. In sharp contrast to this prediction, however, we observe a well-quantized Hall plateau at $(2/3)(h/e^2)$; this is best seen in Fig. 3(c) which zooms in on $R_{xy}$ near $\nu=3/2$. The quantized Hall resistance is also observed if the current and voltage leads are exchanged (see the $R_{yx}$ trace shown in Fig. 3(d)) \cite{Suppl.Mat.}. We are therefore observing an incompressible anisotropic FQHS and not a compressible stripe phase.

The state we observe might be a \textit{striped FQHS} composed of alternating Pfaffian and anti-Pfaffian stripes, as theoretically discussed recently \cite{Wan.PRB.2016}, although it is unclear how to link the anisotropies seen in our data to the anisotropic mass in our 2DES. On the other hand, several key characteristics of our data match those of an anisotropic FQHS proposed by Mulligan, Nayak, and Kachru (MNK) \cite{MNK.PRB.2011} to explain some features of an anisotropic FQHS observed in a GaAs 2DES in the $N=1$ LL at $\nu=7/3$ when a parallel magnetic field ($B_{||}$) is applied \cite{Suppl.Mat., Xia.Natphys.2011}. MNK report that, subtle changes in the electron-electron interaction in the presence of a $B_{||}$ can lead to a \textit{nematic FQHS} with anisotropic transport coefficients below a critical temperature ($T_C$) whose magnitude depends on $B_{||}$ and the details of the sample's structure and transport coefficients. Above $T_C$, the 2DES should exhibit isotropic longitudinal resistances while below $T_C$ the resistance along the hard axis should show an abrupt rise while along the easy axis it should abruptly drop. For $T<T_{C}$, both resistances should drop with decreasing $T$, as expected for a FQHS (see Fig. 1 of MNK). MNK also predict that the sudden transition could be somewhat rounded depending on the experimental details (Fig. 3 of MNK).

\begin{figure*}[t!]
\includegraphics[width=.99\textwidth]{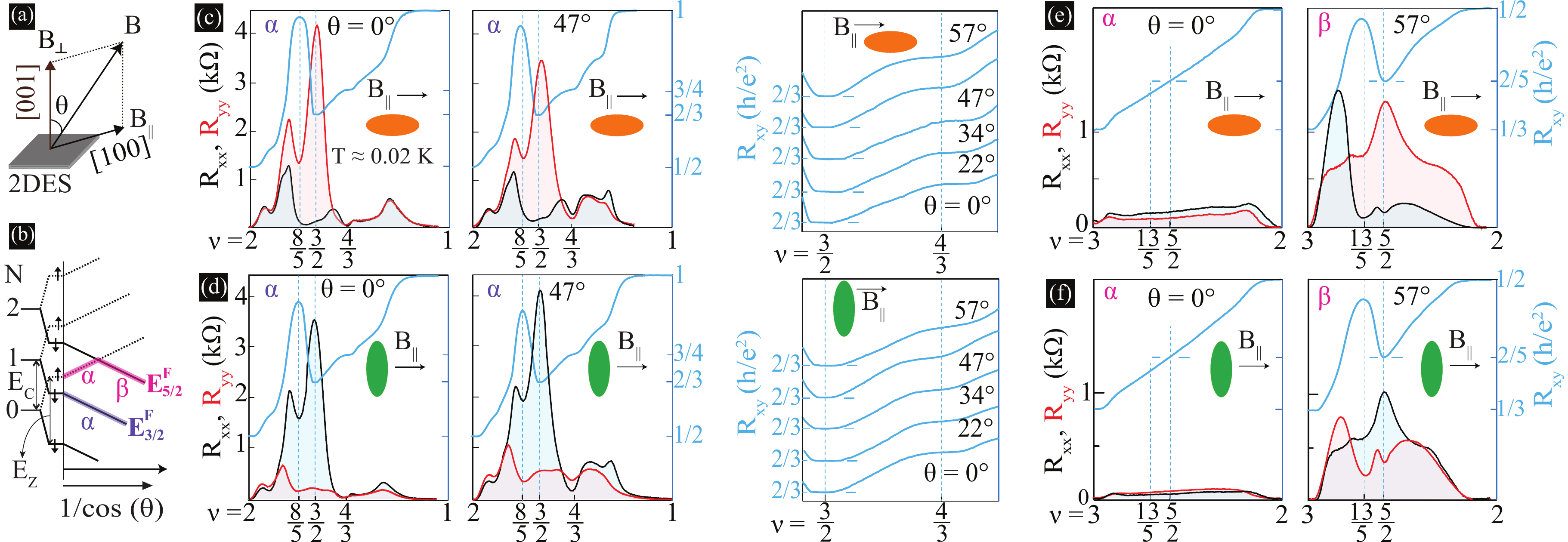}
\caption{\label{fig:Fig4} 
Tilt-dependence of data, all taken at $T\simeq 0.02$ K. (a) The sample is mounted on a rotating stage to support \textit{in situ} tilt. The direction of $B_{||}$ is along [100], i.e., along $R_{xx}$. (b) LL diagram as a function of tilt. Thick purple and pink lines represent $E^F_{\nu=3/2}$ and $E^F_{\nu=5/2}$, respectively. In (c) and (d) data are shown for $2>\nu>1$ and in (e) and (f) for $3>\nu>2$. (c) and (e) contain data for when the 2D electrons occupy X, and (d) and (f) when they are in Y.
}
\end{figure*}

Our \textit{T}-dependent data, summarized in Fig. 3, qualitatively show many of the features predicted by MNK. Considering the low-temperature traces in Fig. 3(a), we associate $R_{xx}$ with the easy axis of the low-temperature nematic phase and $R_{yy}$ with the hard axis. Now, starting from high temperatures, we observe that the anisotropy is reversed at the highest $T$, i.e., $R_{xx}>R_{yy}$ at $\nu=3/2$ (Fig. 3(b)). This is different from the prediction of MNK which states that the 2DES should be isotropic at high $T$, but is reasonable given that our 2DES has an intrinsic (mass) anisotropy in the absence of any interaction (see the $B=0$ panel in Fig. 3(b)). As $T$ is lowered from $\simeq0.33$ K to $\simeq0.20$ K, $R_{xx}$ drops rather suddenly while $R_{yy}$ increases significantly. For lower $T$, both $R_{xx}$ and $R_{yy}$ drop with decreasing $T$, consistent with the characteristics of a FQHS, although we cannot reliably extract an energy gap because of the limited ranges of $T$ and the resistance values. It is clear, however, that the ground state is a FQHS as indicted by the robust quantization of $R_{xy}$ at $(2/3)(h/e^2)$ for $T\lesssim0.2$ (Fig. 3(c)).

We emphasize that the theory of MNK was developed for the $\nu=7/3$ FQHS in the $N=1$ LL in titled magnetic fields, and it is not obvious if it would apply to the \textit{even-denominator} FQHS we observe in the $N=1$ LL of a 2DES with anisotropic effective mass without the application of $B_{||}$. It is worth mentioning, however, that an anisotropic FQHS also develops at $\nu=5/2$ at relatively small tilt angles in the GaAs 2DES $N=1$ LL; see, e.g., \cite{Liu.PRB.2013}, and references therein. This state was also interpreted as a nematic FQHS, although no critical temperature or abrupt changes in resistances were observed \cite{Liu.PRB.2013}. We hope that our results would stimulate new theoretical work on anisotropic even-denominator FQHSs in 2DESs with anisotropic masses and, in particular, whether nematicity in such systems would set in below a critical temperature.

Figures 2 and 3 reveal that in our 2DES another highly anisotropic state emerges near $\nu=8/5$ in the $N=1$ LL. Its low-temperature $R_{xx}$ and $R_{yy}$ have the same anisotropy as the $\nu=3/2$ FQHS, but its $R_{xy}$ rises well above the classical Hall line and approaches the quantized value for an \textit{integer} QHS, namely $(h/e^2)$. Such data closely resemble those for bubble phases seen in GaAs in high LLs away from half fillings \cite{Eisenstein.PRL.2002}. However, the conventional signature of a bubble phase is an $R_{xy}$ quantized at the value for the \textit{nearest} integer QHS. Moreover, the longitudinal resistance are \textit{isotropic}. In contrast, we observe an anisotropic phase near $\nu=8/5$ with an $R_{xy}$ which is approaching $(h/e^2)$ instead of $(1/2)(h/e^2)$, implying an unconventional bubble phase comprised of electrons even though the LL is more than half filled. Moreover, we do not see any signs of a bubble phase when the $N=1$ LL is less than half filled in our AlAs 2DES, suggesting a breakdown of particle-hole symmetry. This conjecture is reasonable as the large effective mass in our 2DES can lead to severe LL mixing, which is known to break particle-hole symmetry \cite{Zhang.PRL.2016}.

Next we present data, taken in tilted magnetic fields where $\theta$ denotes the angle between the total field and its perpendicular component (Fig. 4(a)). The additional $B_{||}$ is a useful tuning knob to study the stripe/nematic phases. In GaAs 2DESs, e.g., $B_{||}$ initially turns the isotropic $\nu=5/2$ FQHS in the $N=1$ LL to an anisotropic FQHS, with the hard axis typically along $B_{||}$ \cite{Liu.PRB.2013, Pan.PRL.1999, Lilly.PRL.1999, Jungwirth.PRB.1999, Rezayi.PRL.2000, Shi.PRB.2016}. At sufficiently large $B_{||}$ there is no quantized $R_{xy}$, and the anisotropic state is believed to be a compressible stripe or nematic phase. In our tilt experiments (Fig. 4) we observe very different phenomena: $B_{||}$ does not change the direction of the anisotropy at $\nu=3/2$, even for $B_{||}=12$ T, the largest $B_{||}$ achieved in our measurements. More remarkably, the Hall plateau at $\nu = 3/2$ stays well developed up to the highest $B_{||}$ (Figs. 4(c) and 4(d), right panels). This is unusual because the $\nu = 5/2$ FQHS in GaAs 2DESs is quite sensitive to $B_{||}$ and its Hall plateau disappears for $B_{||}\gtrsim5$ T \cite{Eisenstein.PRL.1988}. The insensitivity of the direction of the anisotropy to the direction of $B_{||}$ and the persistence of the quantized $R_{xy}$ at very large $B_{||}$ suggest that the robustness of the anisotropic $\nu=3/2$ FQHS in our sample is linked to the mass anisotropy.

In Fig. 4 we also show data near $\nu=5/2$ as a function of $\theta$ for the cases when 2D electrons occupy X (panel (e)) or Y (panel (f)). At $\theta=0$, when $E^F_{\nu=5/2}$ lies in 0$\uparrow$ (see Fig. 4(b) LL diagram), transport near $\nu=5/2$ is anisotropic with the same anisotropy as at $B=0$, i.e., resistance is higher along the larger mass direction. At $\theta=57^o$, however, when $E^F_{\nu = 5/2}$ moves to the 2$\downarrow$ LL, the anisotropy direction changes and becomes similar to the $\nu=3/2$ case (i.e., $R_{xx} < R_{yy}$ for X-occupation and $R_{xx} > R_{yy}$ for Y-occupation). There is, however, no clear signature of a Hall plateau, although $R_{xy}$ assumes a value very close to $(2/5)(h/e^2)$. The fact that the anisotropy direction at $\nu=5/2$ in the $N=2$ LL follows the direction of $\nu=3/2$ anisotropy regardless of the direction of $B_{||}$ again attests to the role of mass anisotropy in dictating the orientation of the anisotropic phases in the excited LLs. Note also in the right panels of Figs. 4(e) and 4(f) the emergence of a bubble-like phase near $\nu=13/5$ in the $N=2$ LL. Similar to the phase near $\nu=8/5$, $R_{xy}$ here is again heading toward an unexpected value, namely $(h/e^2)/2$ instead of $(h/e^2)/3$. 

Our results presented here reveal that, unlike in the $N=0$ LL \cite{Suppl.Mat., Yang.PRB.2012}, effective mass anisotropy strongly affects the ground states in higher LLs and induces exotic anisotropic phases, beyond theoretical expectations. They should stimulate further studies of the even-denominator FQHS and, more generally, the many-body physics, in anisotropic systems.

\begin{acknowledgments}
We acknowledge support through the National Science Foundation (Grants DMR 1709076 and ECCS 1508925) for measurements, and the National Science Foundation (Grant No. MRSEC DMR 1420541), the U.S. Department of Energy Basic Energy Science (Grant No. DE-FG02-00-ER45841), and the Gordon and Betty Moore Foundation (Grant No. GBMF4420 for sample fabrication and characterization. This research is funded in part by QuantEmX grants from Institute for Complex Adaptive Matter and the Gordon and Betty Moore Foundation through Grant No. GBMF5305 to M. S. H., M. K. M., and M. S. A portion of this work was performed at the National High Magnetic Field Laboratory, which is supported by National Science Foundation Cooperative Agreement No. DMR-1644779 and the State of Florida. We thank S. Hannahs, T. Murphy, J. Park, H. Baek, and G. Jones at NHMFL for technical support. We also thank J. K. Jain, S. A. Kivelson, S. L. Sondhi, M. Mulligan, I. Sodemann, M. A. Mueed, and Y. Liu for illuminating discussions.
\end{acknowledgments}

\end{document}